\documentclass[twocolumn,english,prl,showpacs]{revtex4-1}
\usepackage[T1]{fontenc}
\usepackage[latin9]{inputenc}
\usepackage{amssymb}
\usepackage{graphicx}
\usepackage{amsmath,color}
\usepackage{mathrsfs}
\usepackage{float}
\usepackage{bm}
\usepackage{indentfirst}
\usepackage{hyperref}

\newcommand{\red}[1]{{#1}}

\begin{document}

\title{
Direct measurement of topological invariants in optical lattices 
}

\author{Lei Wang$^{1,2}$, Alexey A. Soluyanov$^{1}$ and Matthias Troyer$^{1}$}
 
\affiliation{$^{1}$Theoretische Physik, ETH Zurich, 8093 Zurich, Switzerland}
\affiliation{$^{2}$Beijing National Lab for Condensed Matter Physics and Institute
of Physics, Chinese Academy of Sciences, Beijing 100190, China }

\begin{abstract}
We propose an experimental technique for classifying the topology of band structures realized in optical lattices, based on a generalization of topological charge pumping in quantum Hall systems to cold atoms in optical lattices. Time-of-flight measurement along one spatial direction combined with \emph{in situ} detection along the transverse direction provide a direct measure of the system's Chern number, as we illustrate by calculations for the Hofstadter lattice. Based on an analogy with Wannier functions techniques of topological band theory, the method is very general and also allows the measurement of other topological invariants, such as the $\mathbb{Z}_2$ topological invariant of time-reversal symmetric insulators. 

\end{abstract}

\pacs{ 73.43.-f,  67.85.-d, 03.65.Vf}

\maketitle

Recent advances in experimental techniques have led to realization of synthetic gauge fields~\cite{Lin:2009p15747,Lin:2011p29660, Aidelsburger:2011p67955, JimenezGarcia:2012p56707, Struck:2012p56790} and spin-orbit coupling~\cite{Wang:2012p61874,Cheuk:2012p61844} in cold atomic gases. These new developments allow one to study a variety of topological phases of condensed matter physics using cold gases of neutral atoms trapped in optical lattices. Such topological phases occur in systems whose Hilbert space has a non-trivial topological structure, and they are classified according to the value of a corresponding quantum number, the topological invariant.

Particular examples of such phases in condensed matter include the quantum Hall insulators~\cite{Klitzing:1980p64777} or the quantum anomalous Hall insulators~\cite{Haldane:1988p3868}, where the topological invariant of the Hilbert space is the so-called Chern number~\cite{Thouless:1982p24208}. Some recent experiments~\cite{Aidelsburger:2011p67955, JimenezGarcia:2012p56707, Struck:2012p56790, Tarruell:2012p52125} point towards the possibility of realizing an optical lattice with non-zero Chern number in the near future.  

Once the desired lattice is created the question of experimental verification of the non-trivial topology arises. Unlike condensed matter systems, where a routine measurement of the Hall conductance reveals the Chern number value~\cite{Klitzing:1980p64777}, cold atom systems require a special setup~\cite{Brantut:2012p70665} for transport measurements. Edge states probes of condensed matter experiments also become cumbersome in a cold atom environment since the smooth harmonic potential washes out the edge states associated with the quantum Hall state. This problem can potentially be circumvented by enhancing a weak Bragg signal from the topological edge states by means of specifically tuned  Raman transitions~\cite{Buchhold:2012p58396, Goldman:2012p58384}. Other approaches, which do not have an immediate analogy among solid state experiments, might also allow for the measurement of Chern numbers in optical lattices~\cite{Alba:2011p48844, Zhao:2011p45563, Atala:2012ts, Abanin:2012p70250, Goldman:2012p71551}. However, the quest for a universal method to obtain Chern numbers and other topological invariants directly in a single measurement, avoiding a sophisticated experimental setup or data analysis, remains open.



In this Letter we tackle the problem of detecting topological invariants in optical lattice systems from a very different perspective, drawing an analogy to the theory of electric polarization of crystalline solids~\cite{KingSmith:1993p61262} to suggest a simple and effective method to measure Chern numbers in cold atom systems. We introduce the concept of hybrid time-of-flight (HTOF) images, referring to an \emph{in situ} measurement of the cloud's density in one direction during free expansion in the other. The HTOF reveals the topology of the optical lattice just as hybrid Wannier functions (HWF) do in band theory~\cite{Coh:2009p26093, Marzari:2012p65932}. We illustrate our approach by numerical simulations of a square optical lattice that realizes a Hofstadter model~\cite{Hofstadter:1976p4046}, and discuss how it works in lattices with a more complicated geometry. Our  method does not require the presence of the sharp edge states and is not affected by a soft harmonic trap. It can also be used to detect the $\mathbb{Z}_2$ topological invariant of time-reversal ($\cal{T}$) symmetric topological insulators. 


The modern theory of electric polarization of crystalline solids~\cite{KingSmith:1993p61262, Resta:1994p61451} relates the \red{electronic polarization} $P$ to the geometry of the underlying band structure. For a 1D insulator with a single occupied band $ P$=$\frac{1}{2\pi}\oint_{BZ} {\cal A}(k)dk$, where $k$ is the crystal momentum, ${\cal A}(k)$=$i\langle u_{k}|\partial_k|u_{k}\rangle$ is the Berry connection~\cite{Berry:1984p70668} and  
the $u$s are the lattice-periodic parts of the Bloch functions. 
Alternatively, we will use the fact that the polarization can be written~\cite{KingSmith:1993p61262} as the center of mass of the Wannier function constructed for the occupied band~\cite{Kohn:1959p1285}, which can be defined as an \red{expectation value} of position operator projected onto the occupied state~\cite{Kivelson:1982p26324, Marzari:1997p1458}. 

In two dimensions (2D) an insulating Hamiltonian can be viewed as a fictitious 1D insulator subject to an external parameter $k_x$. Polarization of this 1D insulator can be defined by means of HWF~\cite{Sgiarovello:2001p68653}, which are localized in only one direction retaining Bloch character in the other. 
The polarization at each $k_{x}$ is given by the center of mass of the corresponding HWF~\cite{Marzari:2012p65932}.

The definitions of electronic polarization given above are gauge dependent, meaning that $P$ is defined only modulo a lattice vector. For measurements one has to consider the change in polarization induced by a change in an external parameter~\cite{KingSmith:1993p61262}. In the 2D insulator considered above, $k_x$ plays the role of such a parameter. When $k_x$ is adiabatically changed  by $2\pi$, the change in polarization, {\it i.e.} the shift of the HWF center, is proportional to the Chern number~\cite{KingSmith:1993p61262}. This is a manifestation of topological charge pumping~\cite{Thouless:1983p23000, Niu:1990p70657}, with $k_{x}$ being the adiabatic pumping parameter.

 \begin{figure}[tbp]
\centering
  \includegraphics[width=6cm]{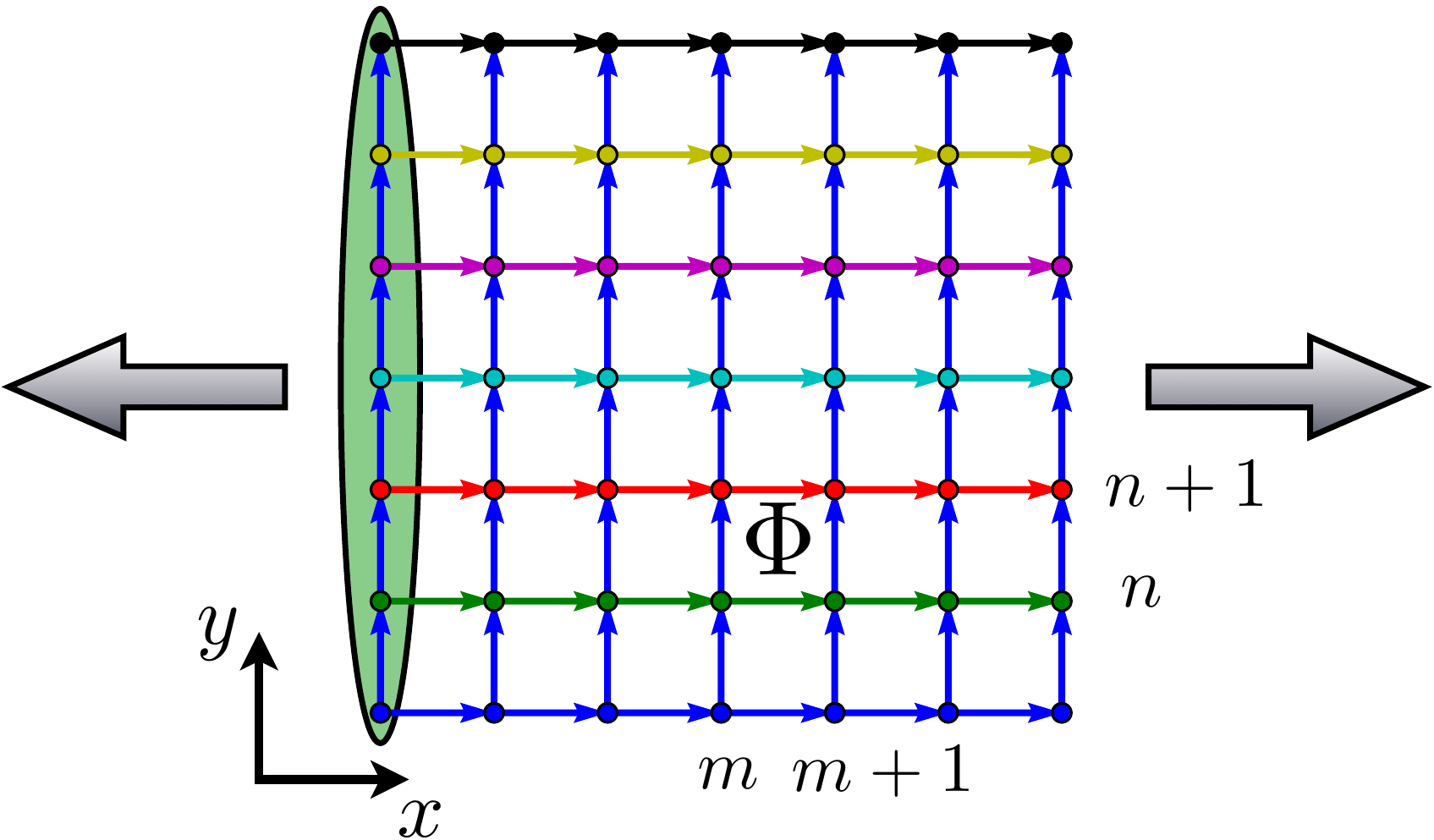}
\caption{Square lattice illustrating the Hofstadter model of Eq.~(\ref{eq:lattice}). The oval marks $q$=$7$ sites of the unit cell. Small arrows indicate the directions in which the phase of the hopping amplitude is chosen to be positive. Different colors of these arrows correspond to different values of the phase. Large arrows indicate the direction of the free expansion of the atomic cloud in hybrid time of flight (HTOF) images.}
\label{fig:lattice}
\end{figure}

A generalization of these ideas to cold atomic gases is a natural way to measure the Chern number in optical lattices. We replace the HWFs of band theory with hybrid densities $\rho(k_{x},y)$, which are the particle densities resolved along the $y$-direction as a function of $k_{x}$. 
Note that, while for an extended system the HWF charge center position can not be reconstructed from the single particle density~\cite{Resta:1994p61451}, this becomes possible in a finite system where the position operator is well defined~\cite{PhysRevLett.80.1800}. 
Hence in a system with finite extent $L_y$ in the $y$-direction we can calculate the HWF center as
\begin{equation}
\bar{y}(k_{x}) = \frac{\int_{0}^{L_y}\, y \rho(k_{x}, y) \mathrm{d} y} {\int_{0}^{L_y}\, \rho(k_{x}, y) \mathrm{d} y}.
\label{eq:continue}
\end{equation}

The proportionality between the shift of the HWF center and the Chern number still holds in the open system. The Chern number measures the charge transported from one boundary to the other as $k_x$ is cycled by $2\pi$.
An experimental measurement of the shift in the hybrid density will hence directly determine the Chern number. 

%
%
%
%
%

Experimentally, $\rho(k_{x},y)$ is measured by an HTOF measurement, in which the lattice and trap are switched off along the $x$-direction, while keeping the lattice and harmonic confinement unchanged in the $y$-direction. In the long time limit~\cite{Gerbier:2008p16148} TOF images map out the crystal momentum distribution along $k_{x}$
\footnote{Due to the short coherence length of the band insulator, the Fresnel interference term~\cite{Gerbier:2008p16148} can safely be ignored for typical expansion times. For example, with $^{40}$K atoms in a typical optical lattice at expansion times of $\sim10$ms will suffice.}. At the same time, the system is still confined in the $y$-direction and a real-space density resolution can be done. 


%
 

We now show HTOF unambiguously determine the Chern number by performing a numerical simulation of the Hofstadter model~\cite{Hofstadter:1976p4046} on a square lattice. Its Hamiltonian is given by
\begin{equation}
H_{\mathrm{lattice}} = -J_x \sum_{m,n} e ^{i2\pi n\Phi}c^{\dagger}_{m+1,n} c_{m,n}- J_y \sum_{m,n}c^{\dagger}_{m,n+1} c_{m,n} + H.c.,
\label{eq:lattice} 
\end{equation}
where $J_\alpha$ is the hopping amplitude in the $\alpha$=$\{x,y\}$ direction and $c_{m,n}$ is the fermionic annihilation operator, with $m$ and $n$ being the column and row indices of the lattice (see Fig~\ref{fig:lattice}). 

An atom hopping clock wise around a plaquette accumulates a phase $\Phi$. We consider $\Phi=p/q$ where $p$ and $q$ are two relatively prime integers. The hopping matrix elements $J_x e ^{i2\pi n\Phi}$ along the $x$-direction depend on the row index $n$ so that each unit cell contains $q$ sites. In the following we fix $q$=$7$ and assume that only the lowest band is occupied. The Chern number $C$ of the lowest band is determined by the Diophantine equation $1$=$qs$+$pC$ \cite{Thouless:1982p24208, Kohmoto:1989p19836, Dana:2000p69285}, where $s$ is an integer and $|C|\le q/2$. 

\begin{figure}[tbp]
\centering
  \includegraphics[width=9cm]{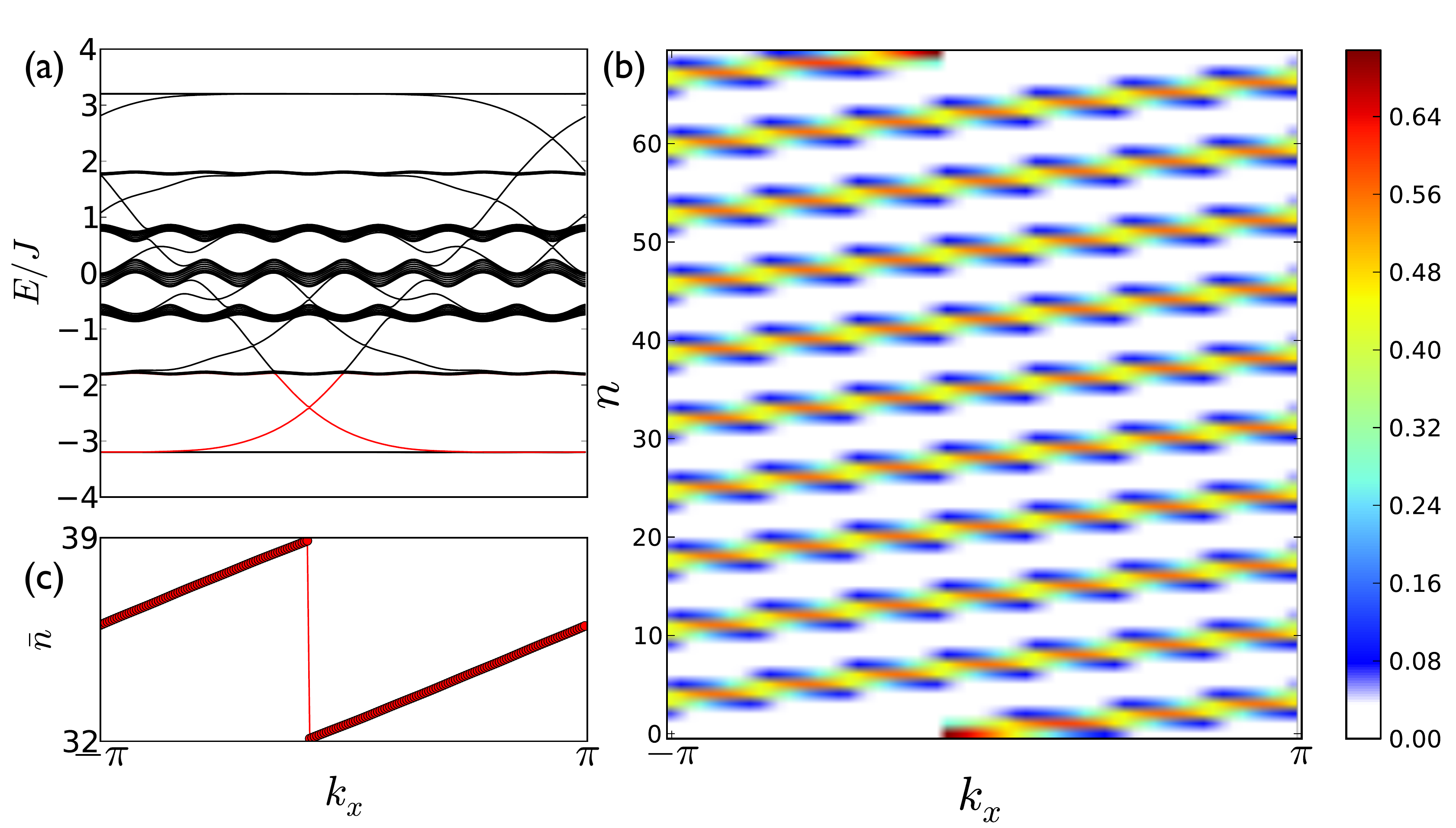}
\caption{Results for a ribbon ($L_{y}=10$) of the Hofstadter lattice model with $p/q=1/7$ and corresponding Chern number $C=1$. (a) Energy spectrum: two edge states (in red) cross the bulk energy gap. (b) Hybrid density $\rho(k_{x},n)$ shifts by one unit cell in a $2\pi$-cycle. (c) The center of mass of the hybrid density Eq.~(\ref{eq:com}) jumps by one unit cell. }
\label{fig:ribbon}
\end{figure}

We first consider an infinite ribbon of this model with width $L_{y}$=$10$ setting $J_x$=$J_y$=$J$, and $p=1$ which corresponds to a Chern number $C=1$. In the  spectrum  shown in Fig.~\ref{fig:ribbon}(a) we see that, as expected for $p=1$, two edge states cross the Fermi level. Analogously to the 2D insulator considered above, this setup can be viewed as a finite 1D chain subject to a $k_{x}$-driven pump. From this point of view the hybrid density $\rho(k_{x}, n)$ describes the change in the density of the 1D system as a function of the pumping parameter. Figure~\ref{fig:ribbon}(b) shows that the hybrid density is shifted by one unit cell in the bulk, indicating that a single charge is pumped across the system, as expected for $C=1$. 

We calculate the $k_{x}$ dependence of the HWF center by taking a tight-binding limit of Eq. (\ref{eq:continue})
%
\begin{equation}
 \bar{n} (k_{x}) = \frac{ \sum_{n} n \rho(k_{x},n)}{ \sum_{n}  \rho(k_{x},n)}.
\label{eq:com}
\end{equation}
As shown in Fig.~\ref{fig:ribbon}(c), $ \bar{n} (k_{x})$ jumps by one unit cell ($q$=$7$ sites), analogous to the HWF shift in Chern insulators~\cite{Coh:2009p26093}.


Having established a clear connection between the hybrid density and Chern number, we now turn to a more realistic case by adding a harmonic trapping potential of the form
\begin{equation}
H_{\mathrm{trap}}=V_{\mathrm{T}} \sum_{m,n} [(m-L_{x}/2)^{2}+(n-qL_{y}/2)^{2}] c^{\dagger}_{m,n} c_{m,n}.  
\label{eq:trap},
\end{equation}
where $L_{x(y)}$ denotes the number of unit cells in the $x(y)$ direction. The system contains $qL_y$ rows and $L_{x}$ columns. The values of $V_\mathrm{T}$ and the number of atoms $N=300$ are chosen such that the atom cloud has a large insulating region corresponding to $1/q$ filling at the trap center. We now consider the cases $p$=$1$ and $p$=$3$, corresponding to Chern numbers $+1$ and $-2$ respectively. 

\begin{figure}[tbp]
\centering
  \includegraphics[width=9cm]{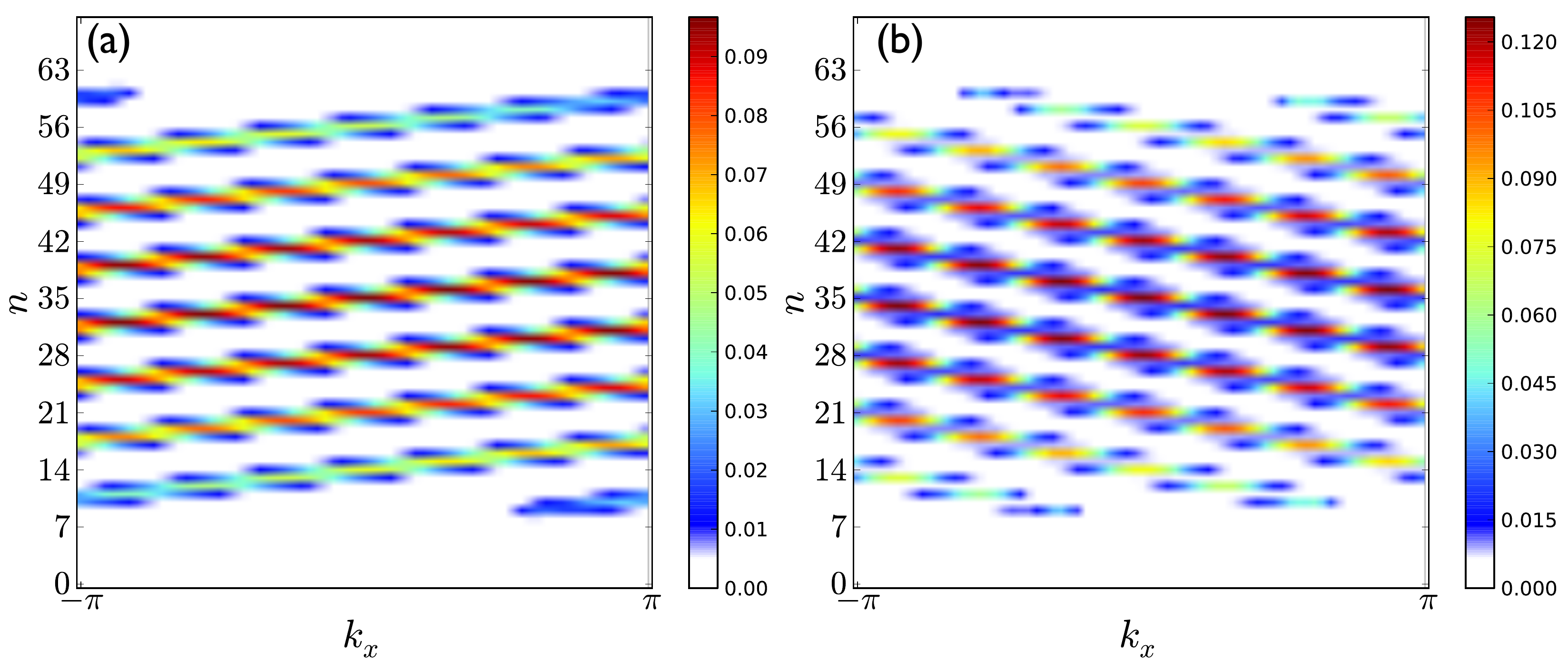}
\caption{The HTOF images for the Hoftadter lattice ($L_{x}=70$, $L_{y}=10$, $N=300$) in the presence of the harmonic trap: (a) for $p/q=1/7$ with $C=1$ ($V_{T}=0.001$), (b) for $p/q=3/7$ with $C=-2$ ($V_{T}=0.001$). Chern numbers can be determined as the number of unit cells traversed by the hybrid density in the course of a $2\pi$-cycle. Upward (downward) direction of the shift corresponds to positive (negative) Chern number. The broadening of lines correspond to exponential localization of the peaks of the hybrid density.}
\label{fig:trap}
\end{figure}

There is no significant difference in the real space densities of the two states since the harmonic trap smears out the edge states~\cite{Buchhold:2012p58396}. In contrast, the  HTOF density profiles allow one to directly read of the Chern numbers. We calculate these HTOF images by  first solving the Schr\"odinger equation $(H_\mathrm{lattice} + H_\mathrm{trap} )\psi_{i}(m,n)= \varepsilon_{i}\psi_{i}(m,n)$ 
and constructing HWF's by means of the Fourier transform in the $x$-direction
\begin{equation}
\psi_{i}(k_{x}, n) =  \sum_{m=0}^{L_{x}-1} e^{i k_{x} m}\psi_{i}(m,n). 
\label{eq:fourier}
\end{equation} 
We then use these wavefunctions to construct the hybrid particle density of the HTOF measurement
\begin{equation}
\rho(k_{x},n) = \frac{1}{N}\sum_{i=1}^{N}|\psi_{i}(k_{x}, n) |^{2}.
\label{eq:rho}
\end{equation}
HTOF images obtained in this way are shown in Fig.~\ref{fig:trap} and clearly exhibit the topological charge pumping effect despite the presence of a trap: the hybrid density shifts by $C$ unit cells along the $y$-direction as $k_{x}$ changes from $-\pi$ to $\pi$. Thus, the hybrid density is an accurate probe of topological properties and allows to directly measure the Chern number. 

\begin{figure}[tbp]
\centering
  \includegraphics[width=9cm]{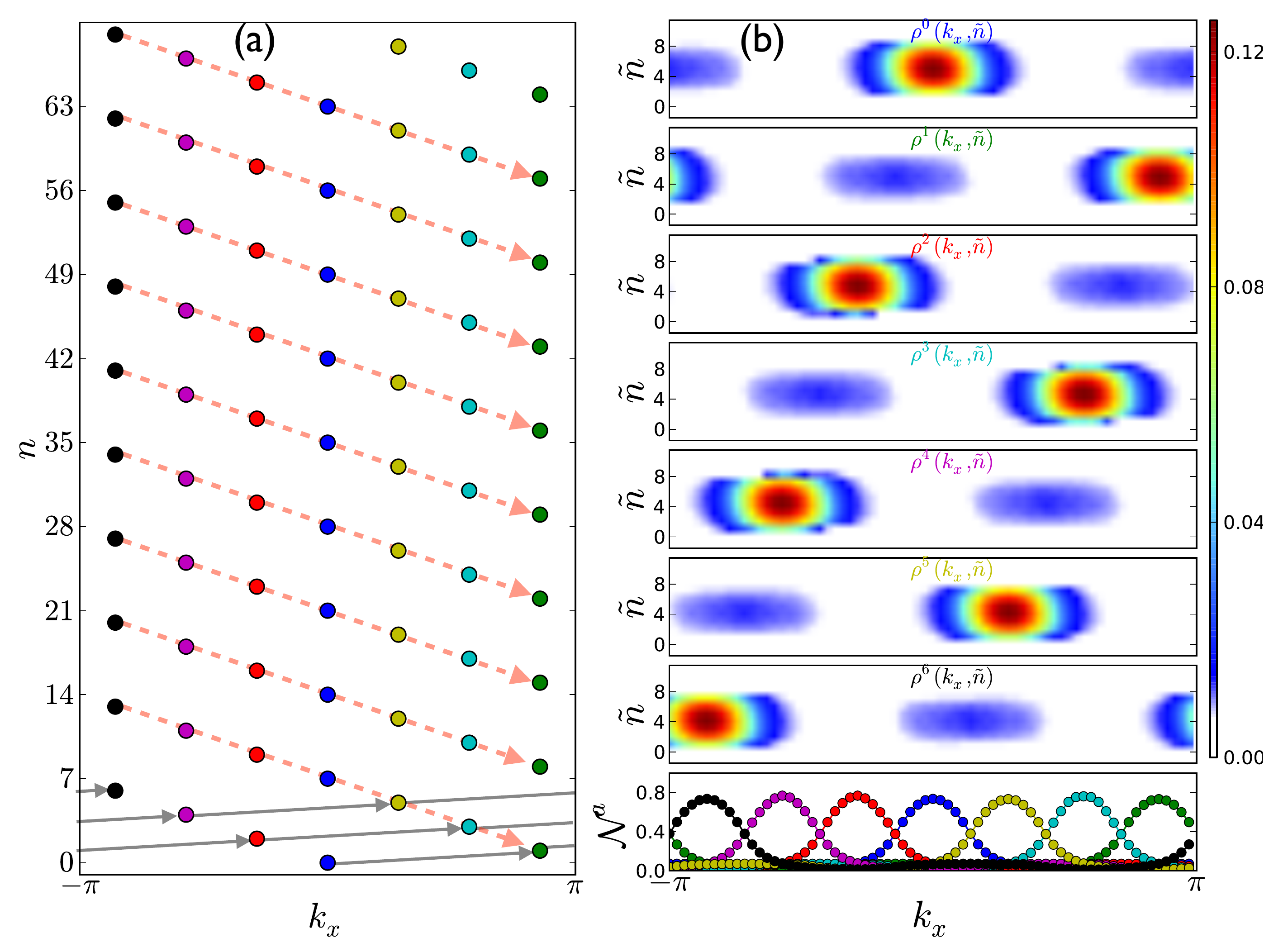}
\caption{(a) Valence band minima of decoupled ($J_y=0$) 1D chains of the Hofstadter model with $ p/q=3/7$. Different colors correspond to different sublattice chains. Solid grey arrows indicate the shift in the dispersion due to the phase factors. Dashed red arrows connect nearest neighbor points and illustrate charge pumping. (b) The sublattice densities $\rho^{a}(k_{x},\tilde{n})$ within a unit cell and the total sublattice density $\mathcal{N}^{a}(k_{x})$ for different values of $a$. The color scheme is the same as in left panel and the model parameters are the same in Fig.~\ref{fig:trap}(b).}
\label{fig:diophantine}
\end{figure}

To get deeper understanding we consider the case of vanishing transverse coupling $J_y$=$0$, corresponding to a set of decoupled tight-binding chains with dispersions $\epsilon_n(k_x)=-2J_x \cos(k_{x} - 2\pi n p/q)$.  The position of the band minimum shifts by $2\pi p/q$ from one chain to the next, as shown in Fig.~\ref{fig:diophantine}(a) for $p$=$3$. If an infinitesimal coupling $J_y$ is added, the 2D lattice is in the Chern insulating regime. Charge pumping can be inferred by tracing the change in the position of the valence band minima for weakly coupled chains: connecting nearest neighbor points in Fig.~\ref{fig:diophantine}(a) reveals a shift by two unit cells ($|C|$=$2$) in the course of a $2\pi$ change of momentum (dashed red arrow in Fig.~\ref{fig:diophantine}(a)). This illustrates the geometrical interpretation of the Diophantine equation~\cite{Thouless:1982p24208}, as also discussed in Ref.~\cite{Huang:2012p70624} in a different context.

The above analysis allows for an alternative way of describing the HTOF results presented in Fig.~\ref{fig:trap}(a-b). We introduce sublattice densities $\rho^{a}(k_{x},\tilde{n})$, which correspond to the particle density on the $a$-th site of the $\tilde{n}$-th unit cell. 
These sublattice densities, shown in  Fig.~\ref{fig:diophantine}(b), shift along the $y$-direction as $k_x$ changes, illustrating the motion of charge. The motion of charge can also be tracked in the total 
sublattice density obtained by summing $\rho^{a}(k_x,\tilde{n})$ over the unit cells:  
\begin{equation}
\mathcal{N}^{a}(k_{x}) = \sum_{\tilde{n}=0}^{L_{y}-1}\rho^{a}(k_{x}, \tilde{n}).
\label{eq:N}
\end{equation}
Thus, the topological nature of the state can also be seen in the total sublattice density, which is potentially accessible in a TOF experiment that can distinguish different sublattices~\cite{Folling:2007p55101, Aidelsburger:2011p67955}. However, such an analysis is specific to the Hofstadter model, while the HTOF measurement is generally applicable.

\begin{figure}[tbp]
\centering
  \includegraphics[width=8cm]{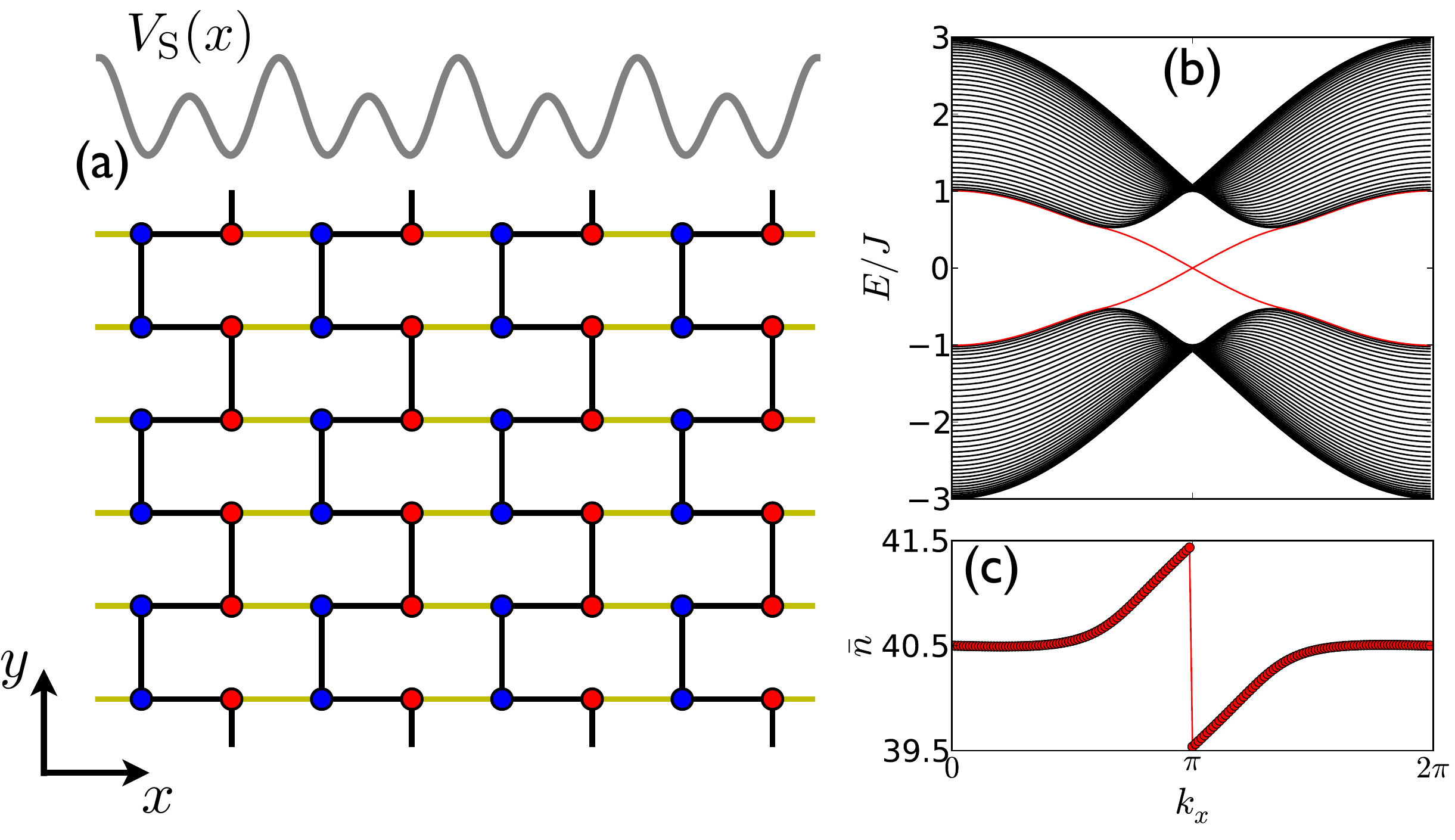}
\caption{(a) The brick-wall lattice split into 1D chains. Thick (black) and light (yellow) bonds indicate intra- and inter-chain hoppings respectively. The shape of an imposed superlattice potential $V_{\rm S}(x)$ is shown schematically (in gray). (b) Energy spectrum of a Haldane model in the ribbon geometry. Edge states are shown in red. (c) The shift of the center of mass of the hybrid density indicates non-trivial Chern number ($C=1$). }
\label{fig:brickwall}
\end{figure}

The square lattice considered so far is particularly simple, since a straightforward partition into 1D chains is possible. HTOF measurement can also work for other lattice geometries such as the honeycomb lattice, which is topologically equivalent to the brick-wall lattice shown in Fig.~\ref{fig:brickwall}(a). Such a lattice was used to create Dirac points in optical lattices~\cite{Tarruell:2012p52125}. This lattice is a potential candidate for realizing  the Haldane model~\cite{Haldane:1988p3868}, which is a canonical example of a Chern insulator. 

A partition of the brick-wall lattice into 1D chains, along which the charge is pumped, is illustrated in Fig.~\ref{fig:brickwall}(a) with solid dark bonds. The chains consist of two sublattices offset from each other in the $x$-direction. Due to this offset the charge pumping is not directly visible in HTOF image unless one separately measures them for each sublattice (for example using the superlattice technique of Ref.~\cite{Folling:2007p55101}).
As illustrated in Fig.~\ref{fig:brickwall}(c) the center of mass of the hybrid density along the zig-zag 1D chain is indeed shifted by one unit cell along the 1D chain, revealing the Chern number $C=1$ of the Haldane model.

%


The HTOF technique can not only determine Chern numbers, but also the $\mathbb{Z}_2$ topological invariant of $\cal{T}$-symmetric insulators~\cite{Hasan:2010p23520}.
In band theory this invariant can be obtained by means of HWFs~\cite{Fu:2006p3991, Soluyanov:2011p33800, Yu:2011p37026}, once they form $\cal{T}$ images of one another. The occupied single particle states of the insulator can always be split into two sets of states (related by $\cal{T}$ symmetry) where each set has a well defined Chern number~\cite{Soluyanov:2012p50967}. For a wide range of models such a splitting can be achieved by projecting the occupied states onto particular spin directions~\cite{Sheng:2006p4513,Prodan:2009p23865}. If the values of thus obtained spin Chern numbers are odd, the system is in the $\mathbb{Z}_2$-insulating regime. In the context of cold atoms, considering a minimal model with only two occupied bands, a spin-projected HTOF measurement of spin-resolved densities would serve as a direct measurement of the spin Chern numbers, and hence of the $\mathbb{Z}_2$-invariant.

The proposed HTOF technique is not only practical, providing exhaustive information about the topological state of an optical lattice, but is also a conceptually novel idea for using a cold atom lattice as a quantum simulator. Hybrid density measurement as proposed here for cold atom systems are not possible in condensed matter experiments and the HWFs are used only in computer simulations. These experiments can thus access novel probes of topological order and will give rise to further implementations of so far numerical experiments of condensed matter in real experiments on cold atom systems.

\paragraph{Acknowledgment} 
We thank David Vanderbilt, Thomas Uehlinger, Daniel Greif and Gregor Jotzu for helpful discussions.
The work was supported by the Swiss National Science Foundation through the NCCR QSIT and the European Research Council.  Simulations were run on the Brutus cluster at ETH Zurich.

\bibliographystyle{apsrev4-1}

\bibliography{hofstadterTOF}

\begin{thebibliography}{45}%
\makeatletter
\providecommand \@ifxundefined [1]{%
 \@ifx{#1\undefined}
}%
\providecommand \@ifnum [1]{%
 \ifnum #1\expandafter \@firstoftwo
 \else \expandafter \@secondoftwo
 \fi
}%
\providecommand \@ifx [1]{%
 \ifx #1\expandafter \@firstoftwo
 \else \expandafter \@secondoftwo
 \fi
}%
\providecommand \natexlab [1]{#1}%
\providecommand \enquote  [1]{``#1''}%
\providecommand \bibnamefont  [1]{#1}%
\providecommand \bibfnamefont [1]{#1}%
\providecommand \citenamefont [1]{#1}%
\providecommand \href@noop [0]{\@secondoftwo}%
\providecommand \href [0]{\begingroup \@sanitize@url \@href}%
\providecommand \@href[1]{\@@startlink{#1}\@@href}%
\providecommand \@@href[1]{\endgroup#1\@@endlink}%
\providecommand \@sanitize@url [0]{\catcode `\\12\catcode `\$12\catcode
  `\&12\catcode `\#12\catcode `\^12\catcode `\_12\catcode `\%12\relax}%
\providecommand \@@startlink[1]{}%
\providecommand \@@endlink[0]{}%
\providecommand \url  [0]{\begingroup\@sanitize@url \@url }%
\providecommand \@url [1]{\endgroup\@href {#1}{\urlprefix }}%
\providecommand \urlprefix  [0]{URL }%
\providecommand \Eprint [0]{\href }%
\@ifxundefined \urlstyle {%
  \providecommand \doi  [0]{\begingroup \@sanitize@url \@doi}%
  \providecommand \@doi [1]{\endgroup \@@startlink {\doibase
  #1}doi:\discretionary {}{}{}#1\@@endlink }%
}{%
  \providecommand \doi  [0]{doi:\discretionary{}{}{}\begingroup
  \urlstyle{rm}\Url }%
}%
\providecommand \doibase [0]{http://dx.doi.org/}%
\providecommand \Doi [0]{\begingroup \@sanitize@url \@Doi }%
\providecommand \@Doi  [1]{\endgroup\@@startlink{\doibase#1}\@@Doi}%
\providecommand \@@Doi [1]{#1\@@endlink}%
\providecommand \selectlanguage [0]{\@gobble}%
\providecommand \bibinfo  [0]{\@secondoftwo}%
\providecommand \bibfield  [0]{\@secondoftwo}%
\providecommand \translation [1]{[#1]}%
\providecommand \BibitemOpen [0]{}%
\providecommand \bibitemStop [0]{}%
\providecommand \bibitemNoStop [0]{.\EOS\space}%
\providecommand \EOS [0]{\spacefactor3000\relax}%
\providecommand \BibitemShut  [1]{\csname bibitem#1\endcsname}%
\bibitem [{\citenamefont {Lin}\ \emph {et~al.}(2009)\citenamefont {Lin},
  \citenamefont {Compton}, \citenamefont {Jim{\'e}nez-Garc{\'\i}a},
  \citenamefont {Porto},\ and\ \citenamefont {Spielman}}]{Lin:2009p15747}%
  \BibitemOpen
  \bibfield  {author} {\bibinfo {author} {\bibfnamefont {Y.~J.}\ \bibnamefont
  {Lin}}, \bibinfo {author} {\bibfnamefont {R.~L.}\ \bibnamefont {Compton}},
  \bibinfo {author} {\bibfnamefont {K.}~\bibnamefont
  {Jim{\'e}nez-Garc{\'\i}a}}, \bibinfo {author} {\bibfnamefont {J.~V.}\
  \bibnamefont {Porto}}, \ and\ \bibinfo {author} {\bibfnamefont {I.~B.}\
  \bibnamefont {Spielman}},\ }\Doi {10.1038/nature08609} {\bibfield  {journal}
  {\bibinfo  {journal} {Nature},\ }\textbf {\bibinfo {volume} {462}},\ \bibinfo
  {pages} {628} (\bibinfo {year} {2009})}\BibitemShut {NoStop}%
\bibitem [{\citenamefont {Lin}\ \emph {et~al.}(2011)\citenamefont {Lin},
  \citenamefont {Compton}, \citenamefont {Jimenez-Garcia}, \citenamefont
  {Phillips}, \citenamefont {Porto},\ and\ \citenamefont
  {Spielman}}]{Lin:2011p29660}%
  \BibitemOpen
  \bibfield  {author} {\bibinfo {author} {\bibfnamefont {Y.-J.}\ \bibnamefont
  {Lin}}, \bibinfo {author} {\bibfnamefont {R.~L.}\ \bibnamefont {Compton}},
  \bibinfo {author} {\bibfnamefont {K.}~\bibnamefont {Jimenez-Garcia}},
  \bibinfo {author} {\bibfnamefont {W.~D.}\ \bibnamefont {Phillips}}, \bibinfo
  {author} {\bibfnamefont {J.~V.}\ \bibnamefont {Porto}}, \ and\ \bibinfo
  {author} {\bibfnamefont {I.~B.}\ \bibnamefont {Spielman}},\ }\Doi
  {10.1038/NPHYS1954} {\bibfield  {journal} {\bibinfo  {journal} {Nat Phys},\
  }\textbf {\bibinfo {volume} {7}},\ \bibinfo {pages} {531} (\bibinfo {year}
  {2011})}\BibitemShut {NoStop}%
\bibitem [{\citenamefont {Aidelsburger}\ \emph {et~al.}(2011)\citenamefont
  {Aidelsburger}, \citenamefont {Atala}, \citenamefont {Nascimb{\`e}ne},
  \citenamefont {Trotzky}, \citenamefont {Chen},\ and\ \citenamefont
  {Bloch}}]{Aidelsburger:2011p67955}%
  \BibitemOpen
  \bibfield  {author} {\bibinfo {author} {\bibfnamefont {M.}~\bibnamefont
  {Aidelsburger}}, \bibinfo {author} {\bibfnamefont {M.}~\bibnamefont {Atala}},
  \bibinfo {author} {\bibfnamefont {S.}~\bibnamefont {Nascimb{\`e}ne}},
  \bibinfo {author} {\bibfnamefont {S.}~\bibnamefont {Trotzky}}, \bibinfo
  {author} {\bibfnamefont {Y.}~\bibnamefont {Chen}}, \ and\ \bibinfo {author}
  {\bibfnamefont {I.}~\bibnamefont {Bloch}},\ }\href@noop {} {\bibfield
  {journal} {\bibinfo  {journal} {Phys. Rev. Lett.},\ }\textbf {\bibinfo
  {volume} {107}},\ \bibinfo {pages} {255301} (\bibinfo {year}
  {2011})}\BibitemShut {NoStop}%
\bibitem [{\citenamefont {Jimenez-Garcia}\ \emph {et~al.}(2012)\citenamefont
  {Jimenez-Garcia}, \citenamefont {LeBlanc}, \citenamefont {Williams},
  \citenamefont {Beeler}, \citenamefont {Perry},\ and\ \citenamefont
  {Spielman}}]{JimenezGarcia:2012p56707}%
  \BibitemOpen
  \bibfield  {author} {\bibinfo {author} {\bibfnamefont {K.}~\bibnamefont
  {Jimenez-Garcia}}, \bibinfo {author} {\bibfnamefont {L.}~\bibnamefont
  {LeBlanc}}, \bibinfo {author} {\bibfnamefont {R.}~\bibnamefont {Williams}},
  \bibinfo {author} {\bibfnamefont {M.}~\bibnamefont {Beeler}}, \bibinfo
  {author} {\bibfnamefont {A.}~\bibnamefont {Perry}}, \ and\ \bibinfo {author}
  {\bibfnamefont {I.}~\bibnamefont {Spielman}},\ }\href@noop {} {\bibfield
  {journal} {\bibinfo  {journal} {Phys. Rev. Lett.},\ }\textbf {\bibinfo
  {volume} {108}},\ \bibinfo {pages} {225303} (\bibinfo {year}
  {2012})}\BibitemShut {NoStop}%
\bibitem [{\citenamefont {Struck}\ \emph {et~al.}(2012)\citenamefont {Struck},
  \citenamefont {{\"O}lschl{\"a}ger}, \citenamefont {Weinberg}, \citenamefont
  {Hauke}, \citenamefont {Simonet}, \citenamefont {Eckardt}, \citenamefont
  {Lewenstein}, \citenamefont {Sengstock},\ and\ \citenamefont
  {Windpassinger}}]{Struck:2012p56790}%
  \BibitemOpen
  \bibfield  {author} {\bibinfo {author} {\bibfnamefont {J.}~\bibnamefont
  {Struck}}, \bibinfo {author} {\bibfnamefont {C.}~\bibnamefont
  {{\"O}lschl{\"a}ger}}, \bibinfo {author} {\bibfnamefont {M.}~\bibnamefont
  {Weinberg}}, \bibinfo {author} {\bibfnamefont {P.}~\bibnamefont {Hauke}},
  \bibinfo {author} {\bibfnamefont {J.}~\bibnamefont {Simonet}}, \bibinfo
  {author} {\bibfnamefont {A.}~\bibnamefont {Eckardt}}, \bibinfo {author}
  {\bibfnamefont {M.}~\bibnamefont {Lewenstein}}, \bibinfo {author}
  {\bibfnamefont {K.}~\bibnamefont {Sengstock}}, \ and\ \bibinfo {author}
  {\bibfnamefont {P.}~\bibnamefont {Windpassinger}},\ }\href@noop {} {\bibfield
   {journal} {\bibinfo  {journal} {Phys. Rev. Lett.},\ }\textbf {\bibinfo
  {volume} {108}},\ \bibinfo {pages} {225304} (\bibinfo {year}
  {2012})}\BibitemShut {NoStop}%
\bibitem [{\citenamefont {Wang}\ \emph {et~al.}(2012)\citenamefont {Wang},
  \citenamefont {Yu}, \citenamefont {Fu}, \citenamefont {Miao}, \citenamefont
  {Huang}, \citenamefont {Chai}, \citenamefont {Zhai},\ and\ \citenamefont
  {Zhang}}]{Wang:2012p61874}%
  \BibitemOpen
  \bibfield  {author} {\bibinfo {author} {\bibfnamefont {P.}~\bibnamefont
  {Wang}}, \bibinfo {author} {\bibfnamefont {Z.-Q.}\ \bibnamefont {Yu}},
  \bibinfo {author} {\bibfnamefont {Z.}~\bibnamefont {Fu}}, \bibinfo {author}
  {\bibfnamefont {J.}~\bibnamefont {Miao}}, \bibinfo {author} {\bibfnamefont
  {L.}~\bibnamefont {Huang}}, \bibinfo {author} {\bibfnamefont
  {S.}~\bibnamefont {Chai}}, \bibinfo {author} {\bibfnamefont {H.}~\bibnamefont
  {Zhai}}, \ and\ \bibinfo {author} {\bibfnamefont {J.}~\bibnamefont {Zhang}},\
  }\href@noop {} {\bibfield  {journal} {\bibinfo  {journal} {Phys. Rev.
  Lett.},\ }\textbf {\bibinfo {volume} {109}},\ \bibinfo {pages} {095301}
  (\bibinfo {year} {2012})}\BibitemShut {NoStop}%
\bibitem [{\citenamefont {Cheuk}\ \emph {et~al.}(2012)\citenamefont {Cheuk},
  \citenamefont {Sommer}, \citenamefont {Hadzibabic}, \citenamefont {Yefsah},
  \citenamefont {Bakr},\ and\ \citenamefont {Zwierlein}}]{Cheuk:2012p61844}%
  \BibitemOpen
  \bibfield  {author} {\bibinfo {author} {\bibfnamefont {L.~W.}\ \bibnamefont
  {Cheuk}}, \bibinfo {author} {\bibfnamefont {A.~T.}\ \bibnamefont {Sommer}},
  \bibinfo {author} {\bibfnamefont {Z.}~\bibnamefont {Hadzibabic}}, \bibinfo
  {author} {\bibfnamefont {T.}~\bibnamefont {Yefsah}}, \bibinfo {author}
  {\bibfnamefont {W.~S.}\ \bibnamefont {Bakr}}, \ and\ \bibinfo {author}
  {\bibfnamefont {M.~W.}\ \bibnamefont {Zwierlein}},\ }\href@noop {} {\bibfield
   {journal} {\bibinfo  {journal} {Phys. Rev. Lett.},\ }\textbf {\bibinfo
  {volume} {109}},\ \bibinfo {pages} {095302} (\bibinfo {year}
  {2012})}\BibitemShut {NoStop}%
\bibitem [{\citenamefont {Klitzing}\ \emph {et~al.}(1980)\citenamefont
  {Klitzing}, \citenamefont {Dorda},\ and\ \citenamefont
  {Pepper}}]{Klitzing:1980p64777}%
  \BibitemOpen
  \bibfield  {author} {\bibinfo {author} {\bibfnamefont {K.}~\bibnamefont
  {Klitzing}}, \bibinfo {author} {\bibfnamefont {G.}~\bibnamefont {Dorda}}, \
  and\ \bibinfo {author} {\bibfnamefont {M.}~\bibnamefont {Pepper}},\ }\Doi
  {10.1103/PhysRevLett.45.494} {\bibfield  {journal} {\bibinfo  {journal}
  {Phys. Rev. Lett.},\ }\textbf {\bibinfo {volume} {45}},\ \bibinfo {pages}
  {494} (\bibinfo {year} {1980})}\BibitemShut {NoStop}%
\bibitem [{\citenamefont {Haldane}(1988)}]{Haldane:1988p3868}%
  \BibitemOpen
  \bibfield  {author} {\bibinfo {author} {\bibfnamefont {F.~D.~M.}\
  \bibnamefont {Haldane}},\ }\href@noop {} {\bibfield  {journal} {\bibinfo
  {journal} {Phys. Rev. Lett.},\ }\textbf {\bibinfo {volume} {61}},\ \bibinfo
  {pages} {2015} (\bibinfo {year} {1988})}\BibitemShut {NoStop}%
\bibitem [{\citenamefont {Thouless}\ \emph {et~al.}(1982)\citenamefont
  {Thouless}, \citenamefont {Kohmoto}, \citenamefont {Nightingale},\ and\
  \citenamefont {den Nijs}}]{Thouless:1982p24208}%
  \BibitemOpen
  \bibfield  {author} {\bibinfo {author} {\bibfnamefont {D.~J.}\ \bibnamefont
  {Thouless}}, \bibinfo {author} {\bibfnamefont {M.}~\bibnamefont {Kohmoto}},
  \bibinfo {author} {\bibfnamefont {M.~P.}\ \bibnamefont {Nightingale}}, \ and\
  \bibinfo {author} {\bibfnamefont {M.}~\bibnamefont {den Nijs}},\ }\href@noop
  {} {\bibfield  {journal} {\bibinfo  {journal} {Phys. Rev. Lett.},\ }\textbf
  {\bibinfo {volume} {49}},\ \bibinfo {pages} {405} (\bibinfo {year}
  {1982})}\BibitemShut {NoStop}%
\bibitem [{\citenamefont {Tarruell}\ \emph {et~al.}(2012)\citenamefont
  {Tarruell}, \citenamefont {Greif}, \citenamefont {Uehlinger}, \citenamefont
  {Jotzu},\ and\ \citenamefont {Esslinger}}]{Tarruell:2012p52125}%
  \BibitemOpen
  \bibfield  {author} {\bibinfo {author} {\bibfnamefont {L.}~\bibnamefont
  {Tarruell}}, \bibinfo {author} {\bibfnamefont {D.}~\bibnamefont {Greif}},
  \bibinfo {author} {\bibfnamefont {T.}~\bibnamefont {Uehlinger}}, \bibinfo
  {author} {\bibfnamefont {G.}~\bibnamefont {Jotzu}}, \ and\ \bibinfo {author}
  {\bibfnamefont {T.}~\bibnamefont {Esslinger}},\ }\Doi {10.1038/nature10871}
  {\bibfield  {journal} {\bibinfo  {journal} {Nature},\ }\textbf {\bibinfo
  {volume} {483}},\ \bibinfo {pages} {302} (\bibinfo {year}
  {2012})}\BibitemShut {NoStop}%
\bibitem [{\citenamefont {Brantut}\ \emph {et~al.}(2012)\citenamefont
  {Brantut}, \citenamefont {Meineke}, \citenamefont {Stadler}, \citenamefont
  {Krinner},\ and\ \citenamefont {Esslinger}}]{Brantut:2012p70665}%
  \BibitemOpen
  \bibfield  {author} {\bibinfo {author} {\bibfnamefont {J.-P.}\ \bibnamefont
  {Brantut}}, \bibinfo {author} {\bibfnamefont {J.}~\bibnamefont {Meineke}},
  \bibinfo {author} {\bibfnamefont {D.}~\bibnamefont {Stadler}}, \bibinfo
  {author} {\bibfnamefont {S.}~\bibnamefont {Krinner}}, \ and\ \bibinfo
  {author} {\bibfnamefont {T.}~\bibnamefont {Esslinger}},\ }\Doi
  {10.1126/science.1223175} {\bibfield  {journal} {\bibinfo  {journal}
  {Science},\ }\textbf {\bibinfo {volume} {337}},\ \bibinfo {pages} {1069}
  (\bibinfo {year} {2012})}\BibitemShut {NoStop}%
\bibitem [{\citenamefont {Buchhold}\ \emph {et~al.}(2012)\citenamefont
  {Buchhold}, \citenamefont {Cocks},\ and\ \citenamefont
  {Hofstetter}}]{Buchhold:2012p58396}%
  \BibitemOpen
  \bibfield  {author} {\bibinfo {author} {\bibfnamefont {M.}~\bibnamefont
  {Buchhold}}, \bibinfo {author} {\bibfnamefont {D.}~\bibnamefont {Cocks}}, \
  and\ \bibinfo {author} {\bibfnamefont {W.}~\bibnamefont {Hofstetter}},\ }\Doi
  {10.1103/PhysRevA.85.063614} {\bibfield  {journal} {\bibinfo  {journal} {Phys
  Rev A},\ }\textbf {\bibinfo {volume} {85}},\ \bibinfo {pages} {063614}
  (\bibinfo {year} {2012})}\BibitemShut {NoStop}%
\bibitem [{\citenamefont {Goldman}\ \emph
  {et~al.}(2012){\natexlab{a}}\citenamefont {Goldman}, \citenamefont
  {Beugnon},\ and\ \citenamefont {Gerbier}}]{Goldman:2012p58384}%
  \BibitemOpen
  \bibfield  {author} {\bibinfo {author} {\bibfnamefont {N.}~\bibnamefont
  {Goldman}}, \bibinfo {author} {\bibfnamefont {J.}~\bibnamefont {Beugnon}}, \
  and\ \bibinfo {author} {\bibfnamefont {F.}~\bibnamefont {Gerbier}},\ }\Doi
  {10.1103/PhysRevLett.108.255303} {\bibfield  {journal} {\bibinfo  {journal}
  {Phys. Rev. Lett.},\ }\textbf {\bibinfo {volume} {108}},\ \bibinfo {pages}
  {255303} (\bibinfo {year} {2012}{\natexlab{a}})}\BibitemShut {NoStop}%
\bibitem [{\citenamefont {Alba}\ \emph {et~al.}(2011)\citenamefont {Alba},
  \citenamefont {Fernandez-Gonzalvo}, \citenamefont {Mur-Petit}, \citenamefont
  {Pachos},\ and\ \citenamefont {Garcia-Ripoll}}]{Alba:2011p48844}%
  \BibitemOpen
  \bibfield  {author} {\bibinfo {author} {\bibfnamefont {E.}~\bibnamefont
  {Alba}}, \bibinfo {author} {\bibfnamefont {X.}~\bibnamefont
  {Fernandez-Gonzalvo}}, \bibinfo {author} {\bibfnamefont {J.}~\bibnamefont
  {Mur-Petit}}, \bibinfo {author} {\bibfnamefont {J.}~\bibnamefont {Pachos}}, \
  and\ \bibinfo {author} {\bibfnamefont {J.}~\bibnamefont {Garcia-Ripoll}},\
  }\Doi {10.1103/PhysRevLett.107.235301} {\bibfield  {journal} {\bibinfo
  {journal} {Phys. Rev. Lett.},\ }\textbf {\bibinfo {volume} {107}},\ \bibinfo
  {pages} {235301} (\bibinfo {year} {2011})}\BibitemShut {NoStop}%
\bibitem [{\citenamefont {Zhao}\ \emph {et~al.}(2011)\citenamefont {Zhao},
  \citenamefont {Bray-Ali}, \citenamefont {Williams}, \citenamefont
  {Spielman},\ and\ \citenamefont {Satija}}]{Zhao:2011p45563}%
  \BibitemOpen
  \bibfield  {author} {\bibinfo {author} {\bibfnamefont {E.}~\bibnamefont
  {Zhao}}, \bibinfo {author} {\bibfnamefont {N.}~\bibnamefont {Bray-Ali}},
  \bibinfo {author} {\bibfnamefont {C.}~\bibnamefont {Williams}}, \bibinfo
  {author} {\bibfnamefont {I.}~\bibnamefont {Spielman}}, \ and\ \bibinfo
  {author} {\bibfnamefont {I.}~\bibnamefont {Satija}},\ }\Doi
  {10.1103/PhysRevA.84.063629} {\bibfield  {journal} {\bibinfo  {journal}
  {Phys. Rev. A},\ }\textbf {\bibinfo {volume} {84}},\ \bibinfo {pages}
  {063629} (\bibinfo {year} {2011})}\BibitemShut {NoStop}%
\bibitem [{\citenamefont {Atala}\ \emph {et~al.}(2012)\citenamefont {Atala},
  \citenamefont {Aidelsburger}, \citenamefont {Barreiro}, \citenamefont
  {Abanin}, \citenamefont {Kitagawa}, \citenamefont {Demler},\ and\
  \citenamefont {Bloch}}]{Atala:2012ts}%
  \BibitemOpen
  \bibfield  {author} {\bibinfo {author} {\bibfnamefont {M.}~\bibnamefont
  {Atala}}, \bibinfo {author} {\bibfnamefont {M.}~\bibnamefont {Aidelsburger}},
  \bibinfo {author} {\bibfnamefont {J.~T.}\ \bibnamefont {Barreiro}}, \bibinfo
  {author} {\bibfnamefont {D.}~\bibnamefont {Abanin}}, \bibinfo {author}
  {\bibfnamefont {T.}~\bibnamefont {Kitagawa}}, \bibinfo {author}
  {\bibfnamefont {E.}~\bibnamefont {Demler}}, \ and\ \bibinfo {author}
  {\bibfnamefont {I.}~\bibnamefont {Bloch}},\ }\href
  {http://arxiv.org/abs/1212.0572v1} {\bibfield  {journal} {\bibinfo  {journal}
  {arXiv},\ }\textbf {\bibinfo {volume} {cond-mat.quant-gas}} (\bibinfo {year}
  {2012})},\ \Eprint {http://arxiv.org/abs/1212.0572} {1212.0572} \BibitemShut
  {NoStop}%
\bibitem [{\citenamefont {Abanin}\ \emph {et~al.}(2012)\citenamefont {Abanin},
  \citenamefont {Kitagawa}, \citenamefont {Bloch},\ and\ \citenamefont
  {Demler}}]{Abanin:2012p70250}%
  \BibitemOpen
  \bibfield  {author} {\bibinfo {author} {\bibfnamefont {D.~A.}\ \bibnamefont
  {Abanin}}, \bibinfo {author} {\bibfnamefont {T.}~\bibnamefont {Kitagawa}},
  \bibinfo {author} {\bibfnamefont {I.}~\bibnamefont {Bloch}}, \ and\ \bibinfo
  {author} {\bibfnamefont {E.}~\bibnamefont {Demler}},\ }\href
  {http://arxiv.org/abs/1212.0562v1} {\bibfield  {journal} {\bibinfo  {journal}
  {arXiv},\ }\textbf {\bibinfo {volume} {cond-mat.quant-gas}} (\bibinfo {year}
  {2012})},\ \Eprint {http://arxiv.org/abs/1212.0562v1} {1212.0562v1}
  \BibitemShut {NoStop}%
\bibitem [{\citenamefont {Goldman}\ \emph
  {et~al.}(2012){\natexlab{b}}\citenamefont {Goldman}, \citenamefont
  {Dalibard}, \citenamefont {Dauphin}, \citenamefont {Gerbier}, \citenamefont
  {Lewenstein}, \citenamefont {Zoller},\ and\ \citenamefont
  {Spielman}}]{Goldman:2012p71551}%
  \BibitemOpen
  \bibfield  {author} {\bibinfo {author} {\bibfnamefont {N.}~\bibnamefont
  {Goldman}}, \bibinfo {author} {\bibfnamefont {J.}~\bibnamefont {Dalibard}},
  \bibinfo {author} {\bibfnamefont {A.}~\bibnamefont {Dauphin}}, \bibinfo
  {author} {\bibfnamefont {F.}~\bibnamefont {Gerbier}}, \bibinfo {author}
  {\bibfnamefont {M.}~\bibnamefont {Lewenstein}}, \bibinfo {author}
  {\bibfnamefont {P.}~\bibnamefont {Zoller}}, \ and\ \bibinfo {author}
  {\bibfnamefont {I.~B.}\ \bibnamefont {Spielman}},\ }\href
  {http://arxiv.org/abs/1212.5093v1} {\bibfield  {journal} {\bibinfo  {journal}
  {arXiv},\ }\textbf {\bibinfo {volume} {cond-mat.quant-gas}} (\bibinfo {year}
  {2012}{\natexlab{b}})},\ \Eprint {http://arxiv.org/abs/1212.5093v1}
  {1212.5093v1} \BibitemShut {NoStop}%
\bibitem [{\citenamefont {King-Smith}\ and\ \citenamefont
  {Vanderbilt}(1993)}]{KingSmith:1993p61262}%
  \BibitemOpen
  \bibfield  {author} {\bibinfo {author} {\bibfnamefont {R.~D.}\ \bibnamefont
  {King-Smith}}\ and\ \bibinfo {author} {\bibfnamefont {D.}~\bibnamefont
  {Vanderbilt}},\ }\href@noop {} {\bibfield  {journal} {\bibinfo  {journal}
  {Physical Review B},\ }\textbf {\bibinfo {volume} {47}},\ \bibinfo {pages}
  {1651} (\bibinfo {year} {1993})}\BibitemShut {NoStop}%
\bibitem [{\citenamefont {Coh}\ and\ \citenamefont
  {Vanderbilt}(2009)}]{Coh:2009p26093}%
  \BibitemOpen
  \bibfield  {author} {\bibinfo {author} {\bibfnamefont {S.}~\bibnamefont
  {Coh}}\ and\ \bibinfo {author} {\bibfnamefont {D.}~\bibnamefont
  {Vanderbilt}},\ }\Doi {10.1103/PhysRevLett.102.107603} {\bibfield  {journal}
  {\bibinfo  {journal} {Phys. Rev. Lett.},\ }\textbf {\bibinfo {volume}
  {102}},\ \bibinfo {pages} {107603} (\bibinfo {year} {2009})}\BibitemShut
  {NoStop}%
\bibitem [{\citenamefont {Marzari}\ \emph {et~al.}(2012)\citenamefont
  {Marzari}, \citenamefont {Mostofi}, \citenamefont {Yates}, \citenamefont
  {Souza},\ and\ \citenamefont {Vanderbilt}}]{Marzari:2012p65932}%
  \BibitemOpen
  \bibfield  {author} {\bibinfo {author} {\bibfnamefont {N.}~\bibnamefont
  {Marzari}}, \bibinfo {author} {\bibfnamefont {A.}~\bibnamefont {Mostofi}},
  \bibinfo {author} {\bibfnamefont {J.~R.}\ \bibnamefont {Yates}}, \bibinfo
  {author} {\bibfnamefont {I.}~\bibnamefont {Souza}}, \ and\ \bibinfo {author}
  {\bibfnamefont {D.}~\bibnamefont {Vanderbilt}},\ }\href@noop {} {\bibfield
  {journal} {\bibinfo  {journal} {Rev Mod Phys},\ }\textbf {\bibinfo {volume}
  {84}},\ \bibinfo {pages} {1419} (\bibinfo {year} {2012})}\BibitemShut
  {NoStop}%
\bibitem [{\citenamefont {Hofstadter}(1976)}]{Hofstadter:1976p4046}%
  \BibitemOpen
  \bibfield  {author} {\bibinfo {author} {\bibfnamefont {D.}~\bibnamefont
  {Hofstadter}},\ }\href {http://link.aps.org/doi/10.1103/PhysRevB.14.2239}
  {\bibfield  {journal} {\bibinfo  {journal} {Phys Rev B},\ }\textbf {\bibinfo
  {volume} {14}},\ \bibinfo {pages} {2239} (\bibinfo {year}
  {1976})}\BibitemShut {NoStop}%
\bibitem [{\citenamefont {Resta}(1994)}]{Resta:1994p61451}%
  \BibitemOpen
  \bibfield  {author} {\bibinfo {author} {\bibfnamefont {R.}~\bibnamefont
  {Resta}},\ }\href@noop {} {\bibfield  {journal} {\bibinfo  {journal} {Rev.
  Mod. Phys.},\ }\textbf {\bibinfo {volume} {66}},\ \bibinfo {pages} {899}
  (\bibinfo {year} {1994})}\BibitemShut {NoStop}%
\bibitem [{\citenamefont {Berry}(1984)}]{Berry:1984p70668}%
  \BibitemOpen
  \bibfield  {author} {\bibinfo {author} {\bibfnamefont {M.~V.}\ \bibnamefont
  {Berry}},\ }\Doi {10.1098/rspa.1984.0023} {\bibfield  {journal} {\bibinfo
  {journal} {Proceedings of the Royal Society A: Mathematical, Physical and
  Engineering Sciences},\ }\textbf {\bibinfo {volume} {392}},\ \bibinfo {pages}
  {45} (\bibinfo {year} {1984})}\BibitemShut {NoStop}%
\bibitem [{\citenamefont {Kohn}(1959)}]{Kohn:1959p1285}%
  \BibitemOpen
  \bibfield  {author} {\bibinfo {author} {\bibfnamefont {W.}~\bibnamefont
  {Kohn}},\ }\href {http://link.aps.org/doi/10.1103/PhysRev.115.809} {\bibfield
   {journal} {\bibinfo  {journal} {Physical Review},\ }\textbf {\bibinfo
  {volume} {115}},\ \bibinfo {pages} {809} (\bibinfo {year}
  {1959})}\BibitemShut {NoStop}%
\bibitem [{\citenamefont {Kivelson}(1982)}]{Kivelson:1982p26324}%
  \BibitemOpen
  \bibfield  {author} {\bibinfo {author} {\bibfnamefont {S.~A.}\ \bibnamefont
  {Kivelson}},\ }\href@noop {} {\bibfield  {journal} {\bibinfo  {journal} {Phys
  Rev B},\ }\textbf {\bibinfo {volume} {26}},\ \bibinfo {pages} {4269}
  (\bibinfo {year} {1982})}\BibitemShut {NoStop}%
\bibitem [{\citenamefont {Marzari}\ and\ \citenamefont
  {Vanderbilt}(1997)}]{Marzari:1997p1458}%
  \BibitemOpen
  \bibfield  {author} {\bibinfo {author} {\bibfnamefont {N.}~\bibnamefont
  {Marzari}}\ and\ \bibinfo {author} {\bibfnamefont {D.}~\bibnamefont
  {Vanderbilt}},\ }\Doi {10.1103/PhysRevB.56.12847} {\bibfield  {journal}
  {\bibinfo  {journal} {Phys Rev B},\ }\textbf {\bibinfo {volume} {56}},\
  \bibinfo {pages} {12847} (\bibinfo {year} {1997})}\BibitemShut {NoStop}%
\bibitem [{\citenamefont {Sgiarovello}\ \emph {et~al.}(2001)\citenamefont
  {Sgiarovello}, \citenamefont {Peressi},\ and\ \citenamefont
  {Resta}}]{Sgiarovello:2001p68653}%
  \BibitemOpen
  \bibfield  {author} {\bibinfo {author} {\bibfnamefont {C.}~\bibnamefont
  {Sgiarovello}}, \bibinfo {author} {\bibfnamefont {M.}~\bibnamefont
  {Peressi}}, \ and\ \bibinfo {author} {\bibfnamefont {R.}~\bibnamefont
  {Resta}},\ }\href@noop {} {\bibfield  {journal} {\bibinfo  {journal}
  {Physical Review B},\ }\textbf {\bibinfo {volume} {64}},\ \bibinfo {pages}
  {115202} (\bibinfo {year} {2001})}\BibitemShut {NoStop}%
\bibitem [{\citenamefont {Thouless}(1983)}]{Thouless:1983p23000}%
  \BibitemOpen
  \bibfield  {author} {\bibinfo {author} {\bibfnamefont {D.~J.}\ \bibnamefont
  {Thouless}},\ }\href@noop {} {\bibfield  {journal} {\bibinfo  {journal} {Phys
  Rev B},\ }\textbf {\bibinfo {volume} {27}},\ \bibinfo {pages} {6083}
  (\bibinfo {year} {1983})}\BibitemShut {NoStop}%
\bibitem [{\citenamefont {Niu}(1990)}]{Niu:1990p70657}%
  \BibitemOpen
  \bibfield  {author} {\bibinfo {author} {\bibfnamefont {Q.}~\bibnamefont
  {Niu}},\ }\href@noop {} {\bibfield  {journal} {\bibinfo  {journal} {Phys Rev
  Lett},\ }\textbf {\bibinfo {volume} {64}},\ \bibinfo {pages} {1812} (\bibinfo
  {year} {1990})}\BibitemShut {NoStop}%
\bibitem [{\citenamefont {Resta}(1998)}]{PhysRevLett.80.1800}%
  \BibitemOpen
  \bibfield  {author} {\bibinfo {author} {\bibfnamefont {R.}~\bibnamefont
  {Resta}},\ }\Doi {10.1103/PhysRevLett.80.1800} {\bibfield  {journal}
  {\bibinfo  {journal} {Phys. Rev. Lett.},\ }\textbf {\bibinfo {volume} {80}},\
  \bibinfo {pages} {1800} (\bibinfo {year} {1998})}\BibitemShut {NoStop}%
\bibitem [{\citenamefont {Gerbier}\ \emph {et~al.}(2008)\citenamefont
  {Gerbier}, \citenamefont {Trotzky}, \citenamefont {F{\"o}lling},
  \citenamefont {Schnorrberger}, \citenamefont {Thompson}, \citenamefont
  {Widera}, \citenamefont {Bloch}, \citenamefont {Pollet}, \citenamefont
  {Troyer}, \citenamefont {Capogrosso-Sansone}, \citenamefont {Prokof'ev},\
  and\ \citenamefont {Svistunov}}]{Gerbier:2008p16148}%
  \BibitemOpen
  \bibfield  {author} {\bibinfo {author} {\bibfnamefont {F.}~\bibnamefont
  {Gerbier}}, \bibinfo {author} {\bibfnamefont {S.}~\bibnamefont {Trotzky}},
  \bibinfo {author} {\bibfnamefont {S.}~\bibnamefont {F{\"o}lling}}, \bibinfo
  {author} {\bibfnamefont {U.}~\bibnamefont {Schnorrberger}}, \bibinfo {author}
  {\bibfnamefont {J.}~\bibnamefont {Thompson}}, \bibinfo {author}
  {\bibfnamefont {A.}~\bibnamefont {Widera}}, \bibinfo {author} {\bibfnamefont
  {I.~F.}\ \bibnamefont {Bloch}}, \bibinfo {author} {\bibfnamefont
  {L.}~\bibnamefont {Pollet}}, \bibinfo {author} {\bibfnamefont
  {M.}~\bibnamefont {Troyer}}, \bibinfo {author} {\bibfnamefont
  {B.}~\bibnamefont {Capogrosso-Sansone}}, \bibinfo {author} {\bibfnamefont
  {N.}~\bibnamefont {Prokof'ev}}, \ and\ \bibinfo {author} {\bibfnamefont
  {B.}~\bibnamefont {Svistunov}},\ }\Doi {10.1103/PhysRevLett.101.155303}
  {\bibfield  {journal} {\bibinfo  {journal} {Phys Rev Lett},\ }\textbf
  {\bibinfo {volume} {101}},\ \bibinfo {pages} {155303} (\bibinfo {year}
  {2008})}\BibitemShut {NoStop}%
\bibitem [{Note1()}]{Note1}%
  \BibitemOpen
  \bibinfo {note} {Due to the short coherence length of the band insulator, the
  Fresnel interference term~\cite {Gerbier:2008p16148} can safely be ignored
  for typical expansion times. For example, with $^{40}$K atoms in a typical
  optical lattice at expansion times of $\sim 10$ms will suffice.}\BibitemShut
  {Stop}%
\bibitem [{\citenamefont {Kohmoto}(1989)}]{Kohmoto:1989p19836}%
  \BibitemOpen
  \bibfield  {author} {\bibinfo {author} {\bibfnamefont {M.}~\bibnamefont
  {Kohmoto}},\ }\href@noop {} {\bibfield  {journal} {\bibinfo  {journal} {Phys
  Rev B},\ }\textbf {\bibinfo {volume} {39}},\ \bibinfo {pages} {11943}
  (\bibinfo {year} {1989})}\BibitemShut {NoStop}%
\bibitem [{\citenamefont {Dana}\ \emph {et~al.}(2000)\citenamefont {Dana},
  \citenamefont {Avron},\ and\ \citenamefont {Zak}}]{Dana:2000p69285}%
  \BibitemOpen
  \bibfield  {author} {\bibinfo {author} {\bibfnamefont {I.}~\bibnamefont
  {Dana}}, \bibinfo {author} {\bibfnamefont {Y.}~\bibnamefont {Avron}}, \ and\
  \bibinfo {author} {\bibfnamefont {J.}~\bibnamefont {Zak}},\ }\href@noop {}
  {\bibfield  {journal} {\bibinfo  {journal} {Journal of Physics C: Solid State
  Physics},\ }\textbf {\bibinfo {volume} {18}},\ \bibinfo {pages} {L679}
  (\bibinfo {year} {2000})}\BibitemShut {NoStop}%
\bibitem [{\citenamefont {Huang}\ and\ \citenamefont
  {Arovas}(2012)}]{Huang:2012p70624}%
  \BibitemOpen
  \bibfield  {author} {\bibinfo {author} {\bibfnamefont {Z.}~\bibnamefont
  {Huang}}\ and\ \bibinfo {author} {\bibfnamefont {D.}~\bibnamefont {Arovas}},\
  }\href {http://arxiv.org/abs/1201.0733} {\bibfield  {journal} {\bibinfo
  {journal} {Phys. Rev. B},\ }\textbf {\bibinfo {volume} {86}},\ \bibinfo
  {pages} {245109} (\bibinfo {year} {2012})}\BibitemShut {NoStop}%
\bibitem [{\citenamefont {F{\"o}lling}\ \emph {et~al.}(2007)\citenamefont
  {F{\"o}lling}, \citenamefont {Trotzky}, \citenamefont {Cheinet},
  \citenamefont {Feld}, \citenamefont {Saers}, \citenamefont {Widera},
  \citenamefont {M{\"u}ller},\ and\ \citenamefont
  {Bloch}}]{Folling:2007p55101}%
  \BibitemOpen
  \bibfield  {author} {\bibinfo {author} {\bibfnamefont {S.}~\bibnamefont
  {F{\"o}lling}}, \bibinfo {author} {\bibfnamefont {S.}~\bibnamefont
  {Trotzky}}, \bibinfo {author} {\bibfnamefont {P.}~\bibnamefont {Cheinet}},
  \bibinfo {author} {\bibfnamefont {M.}~\bibnamefont {Feld}}, \bibinfo {author}
  {\bibfnamefont {R.}~\bibnamefont {Saers}}, \bibinfo {author} {\bibfnamefont
  {A.}~\bibnamefont {Widera}}, \bibinfo {author} {\bibfnamefont
  {T.}~\bibnamefont {M{\"u}ller}}, \ and\ \bibinfo {author} {\bibfnamefont
  {I.~F.}\ \bibnamefont {Bloch}},\ }\Doi {10.1038/nature06112} {\bibfield
  {journal} {\bibinfo  {journal} {Nature},\ }\textbf {\bibinfo {volume}
  {448}},\ \bibinfo {pages} {1029} (\bibinfo {year} {2007})}\BibitemShut
  {NoStop}%
\bibitem [{\citenamefont {Hasan}\ and\ \citenamefont
  {Kane}(2010)}]{Hasan:2010p23520}%
  \BibitemOpen
  \bibfield  {author} {\bibinfo {author} {\bibfnamefont {M.~Z.}\ \bibnamefont
  {Hasan}}\ and\ \bibinfo {author} {\bibfnamefont {C.~L.}\ \bibnamefont
  {Kane}},\ }\Doi {10.1103/RevModPhys.82.3045} {\bibfield  {journal} {\bibinfo
  {journal} {Rev Mod Phys},\ }\textbf {\bibinfo {volume} {82}},\ \bibinfo
  {pages} {3045} (\bibinfo {year} {2010})}\BibitemShut {NoStop}%
\bibitem [{\citenamefont {Fu}\ and\ \citenamefont {Kane}(2006)}]{Fu:2006p3991}%
  \BibitemOpen
  \bibfield  {author} {\bibinfo {author} {\bibfnamefont {L.}~\bibnamefont
  {Fu}}\ and\ \bibinfo {author} {\bibfnamefont {C.~L.}\ \bibnamefont {Kane}},\
  }\Doi {10.1103/PhysRevB.74.195312} {\bibfield  {journal} {\bibinfo  {journal}
  {Phys Rev B},\ }\textbf {\bibinfo {volume} {74}},\ \bibinfo {pages} {195312}
  (\bibinfo {year} {2006})}\BibitemShut {NoStop}%
\bibitem [{\citenamefont {Soluyanov}\ and\ \citenamefont
  {Vanderbilt}(2011)}]{Soluyanov:2011p33800}%
  \BibitemOpen
  \bibfield  {author} {\bibinfo {author} {\bibfnamefont {A.~A.}\ \bibnamefont
  {Soluyanov}}\ and\ \bibinfo {author} {\bibfnamefont {D.}~\bibnamefont
  {Vanderbilt}},\ }\Doi {10.1103/PhysRevB.83.235401} {\bibfield  {journal}
  {\bibinfo  {journal} {Phys Rev B},\ }\textbf {\bibinfo {volume} {83}},\
  \bibinfo {pages} {235401} (\bibinfo {year} {2011})}\BibitemShut {NoStop}%
\bibitem [{\citenamefont {Yu}\ \emph {et~al.}(2011)\citenamefont {Yu},
  \citenamefont {Qi}, \citenamefont {Bernevig}, \citenamefont {Fang},\ and\
  \citenamefont {Dai}}]{Yu:2011p37026}%
  \BibitemOpen
  \bibfield  {author} {\bibinfo {author} {\bibfnamefont {R.}~\bibnamefont
  {Yu}}, \bibinfo {author} {\bibfnamefont {X.-L.}\ \bibnamefont {Qi}}, \bibinfo
  {author} {\bibfnamefont {B.~A.}\ \bibnamefont {Bernevig}}, \bibinfo {author}
  {\bibfnamefont {Z.}~\bibnamefont {Fang}}, \ and\ \bibinfo {author}
  {\bibfnamefont {X.}~\bibnamefont {Dai}},\ }\Doi {10.1103/PhysRevB.84.075119}
  {\bibfield  {journal} {\bibinfo  {journal} {Phys Rev B},\ }\textbf {\bibinfo
  {volume} {84}},\ \bibinfo {pages} {075119} (\bibinfo {year}
  {2011})}\BibitemShut {NoStop}%
\bibitem [{\citenamefont {Soluyanov}\ and\ \citenamefont
  {Vanderbilt}(2012)}]{Soluyanov:2012p50967}%
  \BibitemOpen
  \bibfield  {author} {\bibinfo {author} {\bibfnamefont {A.}~\bibnamefont
  {Soluyanov}}\ and\ \bibinfo {author} {\bibfnamefont {D.}~\bibnamefont
  {Vanderbilt}},\ }\Doi {10.1103/PhysRevB.85.115415} {\bibfield  {journal}
  {\bibinfo  {journal} {Physical Review B},\ }\textbf {\bibinfo {volume}
  {85}},\ \bibinfo {pages} {115415} (\bibinfo {year} {2012})}\BibitemShut
  {NoStop}%
\bibitem [{\citenamefont {Sheng}\ \emph {et~al.}(2006)\citenamefont {Sheng},
  \citenamefont {Weng}, \citenamefont {Sheng},\ and\ \citenamefont
  {Haldane}}]{Sheng:2006p4513}%
  \BibitemOpen
  \bibfield  {author} {\bibinfo {author} {\bibfnamefont {D.}~\bibnamefont
  {Sheng}}, \bibinfo {author} {\bibfnamefont {Z.-Y.}\ \bibnamefont {Weng}},
  \bibinfo {author} {\bibfnamefont {L.}~\bibnamefont {Sheng}}, \ and\ \bibinfo
  {author} {\bibfnamefont {F.~D.~M.}\ \bibnamefont {Haldane}},\ }\Doi
  {10.1103/PhysRevLett.97.036808} {\bibfield  {journal} {\bibinfo  {journal}
  {Phys Rev Lett},\ }\textbf {\bibinfo {volume} {97}},\ \bibinfo {pages}
  {036808} (\bibinfo {year} {2006})}\BibitemShut {NoStop}%
\bibitem [{\citenamefont {Prodan}(2009)}]{Prodan:2009p23865}%
  \BibitemOpen
  \bibfield  {author} {\bibinfo {author} {\bibfnamefont {E.}~\bibnamefont
  {Prodan}},\ }\Doi {10.1103/PhysRevB.80.125327} {\bibfield  {journal}
  {\bibinfo  {journal} {Phys Rev B},\ }\textbf {\bibinfo {volume} {80}},\
  \bibinfo {pages} {125327} (\bibinfo {year} {2009})}\BibitemShut {NoStop}%
\end{thebibliography}%

\end{document}